\documentclass[trackchanges, twocolumn]{aastex701}
\usepackage{bm}
\usepackage{booktabs}
\usepackage{graphicx}
\usepackage{subcaption}

\shorttitle{X-ray Jets at z $>$ 2}
\shortauthors{Bettwy et al.}
\begin{document}

\title{X-Ray Observations of Four \emph{z}\,$>$\,2 Radio-loud AGNs with Extended kpc-scale Relativistic Jets}

%\correspondingauthor{Kyra Bettwy} 
\author[orcid=0009-0007-8010-8562,gname=Kyra,sname=Bettwy]{Kyra Bettwy}
\email{kyra.bettwy@yale.edu}
\affiliation{Center for Astrophysics $\vert$\ Harvard\ \&\ Smithsonian, 60 Garden St., Cambridge, MA 02138, USA}
\affiliation{Department of Astronomy, Yale University, 219 Prospect Street, New Haven, CT 06511, USA}
\author[orcid=0000-0003-1516-9450,gname=Luca,sname=Ighina]{Luca Ighina}
\email{luca.ighina@cfa.harvard.edu}
\affiliation{Center for Astrophysics $\vert$\ Harvard\ \&\ Smithsonian, 60 Garden St., Cambridge, MA 02138, USA}
\author[orcid=0000-0002-7898-7664,gname=Thomas,sname=Connor]{Thomas Connor}
\email{thomas.connor@cfa.harvard.edu}
\affiliation{Center for Astrophysics $\vert$\ Harvard\ \&\ Smithsonian, 60 Garden St., Cambridge, MA 02138, USA}

\author[0000-0002-1727-1224]{Emmanuel K. Bempong-Manful}
\email{emmanuel.bempong-manful@manchester.ac.uk}
\affiliation{Jodrell Bank Centre for Astrophysics, Department of Physics and Astronomy, The University of Manchester, Manchester M13 9PL, UK}
\affiliation{School of Physics, University of Bristol, Tyndall Avenue, Bristol BS8 1TL, UK}
\author[0000-0002-9770-0315]{Alberto Moretti}
\email{alberto.moretti@inaf.it}
\affiliation{INAF, Osservatorio Astronomico di Brera, via Brera 28, 20121, Milano, Italy}
\author[0000-0002-0905-7375]{Aneta Siemiginowska}
\email{asiemiginowska@cfa.harvard.edu}
\affil{Center for Astrophysics $\vert$\ Harvard\ \&\ Smithsonian, 60 Garden St., Cambridge, MA 02138, USA}
\author[0000-0002-2231-6861]{Cristiana Spingola}
\email{spingola@ira.inaf.it}
\affiliation{INAF $-$ Istituto di Radioastronomia, Via Gobetti 101, I$-$40129, Bologna, Italy}

\begin{abstract}
At high redshift, the increasing cosmic microwave background (CMB) energy density can enhance kpc-scale X-ray jet emission, requiring both radio and X-ray observations to relate jet properties to physical feedback. In this work we present the analysis of Chandra X-ray observations of four radio-loud AGNs at $z \sim 2\text{--}3$, selected to host kpc-scale radio jets aligned with the pc-scale jets seen in VLBI observations. Although these sources were not selected on their X-ray properties, we find extended X-ray emission on kpc scales for three of the four sources, detected along the direction of both the VLBI-detected jets and extended emission in the lower-resolution VLA imaging. We use the Chandra X-ray data to constrain jet fluxes and luminosities and, combined with complementary radio data, compare the X-ray and radio emission. The discovery of these three jets at X-ray energies, along with one candidate jet and one candidate counterjet,  expands the number of X-ray–detected jets at $z > 2$ to twenty-one in total, enabling not just individual analysis of these sources but contributing to a growing sample for future studies of high-redshift jet emission.

\end{abstract}


\section{Introduction}

The study of active galactic nuclei (AGNs) is invaluable for understanding the evolution and growth rate of supermassive black holes (SMBHs) in the early Universe. The known population of higher-redshift quasars has surged in recent years through large scale surveys, e.g., \citet{Wang_2019}; \citet{Yang_2023}; \citet{Banados_2023}; \citet{Belladitta_2025}; \citet{Ighina_2025}; \citet{Mkrtchyan_2026}; \citet{Yang_2026}. Along with accreting matter, a small fraction (10--15$\%$) of AGNs expel matter at relativistic speeds in the form of bipolar jets \citep{Ghisellini_2013}. These jets are created when the accretion disk is highly magnetized, and the jet is collimated through magnetic stress from disk winds \citep{Blandford_2019}. AGNs with bipolar jets are typically observed to be radio-loud, with $R=S_{\rm 5GHz}/S_{\rm4400A}$ greater than 10 \citep{Kellerman_1989}. These jets can have a significant impact on the host galaxy evolution across cosmic time. They can transfer energy into the surrounding gas, resulting in star formation quenching or even enhancement \citep{Blandford_2019}. Observing these jets in the X-ray band and characterizing their spectra \citep[e.g.][]{Liniewicz_2026} is invaluable to understanding their physics and kinematics, particularly at high redshift values \citep{Connor_2021}.

\begin{figure*}[ht!]
\centering
\includegraphics[width=\textwidth]{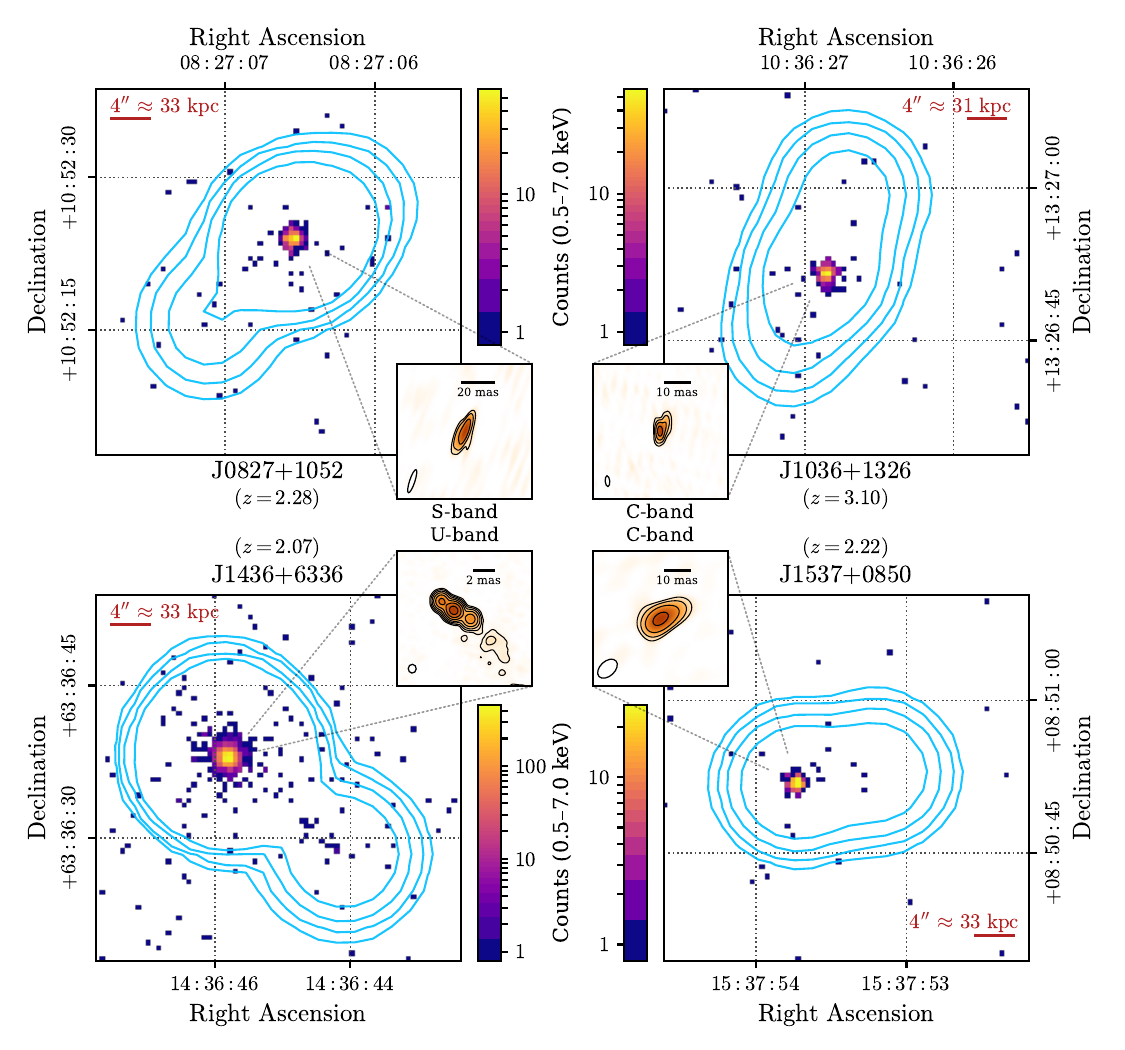}
\caption{Combined Chandra (0.5--7.0) and VLA FIRST (1.4, GHz) observations of the four AGNs. Bumpouts show the pc-scale radio emission seen by the VLBA. FIRST contours are shown at 1, 2, 4 and 8\,mJy\,beam$^{-1}$. VLBA contours start at $5\sigma$ and increase by a factor of 3. Radio observations show consistent jet direction from parsec to kiloparsec scales.}
\label{fig:fig1_images_all}
\end{figure*}

%comment

\begin{deluxetable*}{cccccccccc}
\tabletypesize{\footnotesize}
\tablecaption{Chandra observations for target sources, including individual ObsIDs. Angular wedge annulus parameters listed by source (outer radius set uniformly at 15\arcsec). Redshift values are from \citet{Aghakhanloo_2026}. \label{tab:wedge_params}}
\tablehead{
\colhead{Target} & \colhead{RA} & \colhead{Dec} & \colhead{$z$} & \colhead{$r_\mathrm{in}$} & \colhead{Sectors} & \colhead{Angle} & \colhead{Obs ID} & \colhead{Obs Date} & \colhead{Exp.} \\
\colhead{} & \colhead{(J2000)} & \colhead{(J2000)} & \colhead{} & \colhead{} & \colhead{} & \colhead{} & \colhead{} & & \colhead{(ks)}
}
\startdata
J1036+1326 & 10 36 26.89 & +13 26 51.74 & 3.10 & $3''$ & 8  & $239^\circ$ & 29941 & 2025 Jun 08 & 19.81 \\
J0827+1052 & 08 27 06.51 & +10 52 24.13 & 2.28 & $3''$ & 8  & $215^\circ$ & 29942 & 2025 Jan 06 & 19.57 \\
J1537+0850 & 15 37 53.73 & +08 50 51.88 & 2.22 & $2''$ & 8  & $8^\circ$ & 29943 & 2024 Dec 24 & 9.95  \\
\nodata & \nodata & \nodata & \nodata & \nodata & \nodata & \nodata & 30702 & 2024 Dec 25 & 8.96 \\
J1436+6336 & 14 36 45.80 & +63 36 37.87 & 2.07 & $4''$ & 10 & $320^\circ$ & 29944 & 2025 May 08 & 35.10 \\
\nodata & \nodata & \nodata & \nodata & \nodata & \nodata & \nodata & 30212 & 2026 Jan 05 & 18.27 \\
\enddata
\end{deluxetable*}

Relativistic jet electrons lose energy through two main processes: synchrotron emission and Inverse Compton (IC) scattering. In the radio band, the main emission mechanism in relativistic jets is the synchrotron process. On sub-parsec scales, high-energy emission is created from IC scattering of photons from other AGN components, such as the accretion disk and broad-line region \citep{Ghisellini_2009}. At kiloparsec scales, however, the very first Chandra observations measured unexpectedly bright jets in the X-ray band \citep{Schwartz_2000}, whose emission cannot be explained as an extension of the radio synchrotron spectrum. One proposed explanation is the Inverse Compton scattering of cosmic microwave background photons (IC/CMB) at the kpc scale \citep{Tavecchio_2000}. However, at low redshift the IC model requires large bulk Lorentz factors of 10--20 and an extreme kinetic jet power exceeding $10^{47}$ erg s$^{-1}$ \citep{Tavecchio_2000}. IC/CMB also predicts steady gamma-ray emission from the same electron population, and Fermi-LAT upper limits fall below this predicted level for several low-redshift jets, disfavoring IC/CMB in these sources \citep{Breiding_2023}.

As the CMB energy density scales by a factor of $U_{\rm CMB} \propto (1+z)^4$, IC/CMB is able to dominate energy losses at distances far from the host galaxy \citep{Schwartz_2019}. At low to mid redshift values, however, the dominant emission mechanism remains an open debate: an analysis of the kpc-scale jet of $z = 0.651$ quasar PKS 0637--752 showed that IC/CMB was not a viable option \citep{Meyer_2015, Meyer_2017}, whereas recent inner jet analysis has found that IC/CMB can be consistent with gamma ray upper limits \citep{Maithil_2026}. These results suggest that the dominant emission mechanism not only depends on several physical parameters (e.g. redshift and orientation; \citealt{Meyer_2019}), but can also differ in different regions of the same source. High redshift observations in the X-ray band can reveal structure missed with radio observations, as demonstrated by the discovery of an X-ray jet lacking an obvious radio counterpart \citep{Simionescu_2016}. Yet whether radio non-detections can be informative, and whether X-ray observations should be pursued, depends on quantifying the relative strengths of X-ray and radio emission at high redshift when both are available.

In this work, we present Chandra X-ray ACIS-S observations of four kiloparsec-scale jets in radio-loud quasars at $z>2$: SDSS J143645.80+633637.9 (hereafter J1436+6336), SDSS J153753.72+085051.8 (hereafter J1537+0850), SDSS J082706.51+105224.2 (hereafter J0827+1052), and SDSS J103626.88+132651.7 (hereafter J1036+1326). These are the first X-ray observations of these sources. We constrain the X-ray intensity, spectral shape, and morphology of each jet to work towards constraining the dominant emission mechanism at mid redshift. In Section~\ref{sec:data}, we describe the sample selection and detail the astrometric corrections applied to observations with several exposure times. In Section~\ref{sec:imaging}, we model the point spread function (PSF) of each quasar core and use a Poisson survival function with angular wedge regions to identify potential extended X-ray emission. In Section~\ref{sec:spectral}, we detail spectral analyses of both the core and the extended emission using specextract. In Section~\ref{sec:discussion}, we show that the X-ray and radio measurements of J1436+6336 can be reasonably reproduced by IC/CMB emission, and discuss improvements that future observations can provide. Throughout this paper, we assume a flat $\Lambda$CDM cosmology with $H_0 = 70$~km~s$^{-1}$~Mpc$^{-1}$, $\Omega_{\rm m} = 0.3$, and $\Omega_\Lambda = 0.7$.

\section{Data}
\label{sec:data}

\subsection{Sample Selection and Chandra observations}

The aim of this work is to study the multi-wavelength properties of jets at mid redshifts ($2 \lesssim z \lesssim 3.1$), the range at which the high-energy emission mechanism is expected to transition from being dominated by the synchrotron process to the IC/CMB \citep[e.g.,][]{McKeough_2016, Ighina_2022}. We started our investigation by selecting radio sources in the CLASS radio survey \citep{Myers_2003} and kept objects with SDSS redshifts between $2 \lesssim z \lesssim 3.1$ \citep{Aghakhanloo_2026}. As CLASS is a sample of flat-spectrum radio sources ($\alpha_{\rm r}<0.5$), this sample is biased towards blazars, or jets that point close to our line of sight. We then cross-matched these sources with the VLA FIRST \citep{White_1997} survey to identify sources with extended jet emission. At this step, we also visually screened our sample to reject sources contaminated by potential fore- and background sources. Our final filtering step was to search for extended milliarcsecond-scale emission in archival VLBI observations. We kept only objects that had an additional radio component not associated with any optical/NIR counterpart and that were aligned with the radio jet detected on milliarcsecond scales. From this search, we identified four AGN for further study, which have both parsec-scale radio elongation and larger radio-extensions on scales of 8--15$\arcsec$, as shown in Figure~\ref{fig:fig1_images_all}.

We observed these four sources with the Advanced CCD Imaging Spectrometer \citep[ACIS;][]{Garmire_2003} onboard Chandra through program 26700199 (PI: Ighina); the full set of observations are contained in the Chandra Data Collection (CDC)~\dataset[CDC.730]{https://doi.org/10.25574/cdc.730}. Observations were taken in the VFAINT telemetry mode, with ObsIDs and individual exposure times detailed in Table~\ref{tab:wedge_params}. Each observation was reprocessed using the CIAO \citep[v4.18;][]{Fruscione_2006, Fruscione_2026} script \texttt{chandra\_repro} with CALDB v4.12.4, and images were restricted to the 0.5--7.0 keV band for all subsequent analysis. J1537+0850 and J1436+6336 were each observed in two separate epochs, which were merged for subsequent imaging analysis after applying astrometric corrections as described in Section~\ref{sec:astrometry}.

 The parameters derived from the AGN supplementary radio data are detailed in Table~\ref{tab:combined_results}. The FIRST radio contours are overlaid on the X-ray photon count images, and the corresponding pc-scale VLBA images are shown in Figure~\ref{fig:fig1_images_all}. The VLBA images were accessed through the VLBA calibrator archive\footnote{\url{https://obs.vlba.nrao.edu/cst/}}, associated with the Radio Fundamental Catalogue (RFC\footnote{\url{http://astrogeo.org/rfc/}}; \citealt{Petrov_2025}). The VLBI map for J1436+6336 is an image from the MOJAVE survey \citep{Lister_2009}.

\subsection{Astrometric Correction}
\label{sec:astrometry}

As the focus of this work was to search for extended X-ray emission, misalignments between individual observations of the same target could produce a false signal of extension. We therefore reprocessed the aspect solutions for the two sources with multiple observations in order to ensure minimal deviations in relative astrometry. For each observation, we detected point sources using the CIAO task \texttt{wavdetect} run on wavelet scales of 1.0, 2.0, 4.0, and 8.0 pixels; the resulting positions were crossmatched with external catalogs. For J1537+0850, crossmatching with the Gaia eDR3 catalog yielded 6 matches in the first exposure and 8 in the second. The \texttt{wcs\_match} and \texttt{reproject\_aspect} functions were then used to create a frame transformation and align each observation to the reference catalog, and the corrected event files were merged using \texttt{merge\_obs}. For J1436+6336, crossmatching with the unWISE \citep{Schlafly_2019} and Gaia eDR3 \citep{Smart_2021} catalogs yielded 2 matches in the first exposure and 6 in the second, and the package \texttt{fine\_astro} was used to realign the files. The combined exposure times are provided in Table \ref{tab:wedge_params}. The merged event files are used for imaging analysis only; spectra were extracted from individual observations and combined using \texttt{combine\_spectra} before fitting.

\begin{figure}[t]
\centering
    \includegraphics[width=\columnwidth]{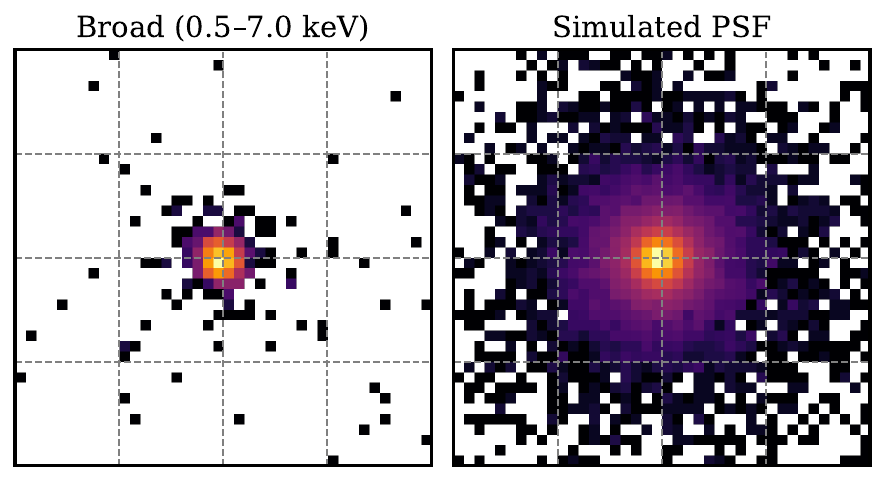}
    \caption{Observed counts and simulated PSF data for J1436+6336, ObsID 29944. Each panel is 20\arcsec by 20\arcsec with dashed lines every 5\arcsec. PSF values were modeled using the software packages ChaRT and MARX; in the field shown, photon counts are 220 times higher in the PSF simulation.}
    \label{fig:real_vs_psf}
\end{figure}

\begin{figure*}[ht!]
\includegraphics[width=0.95\textwidth]{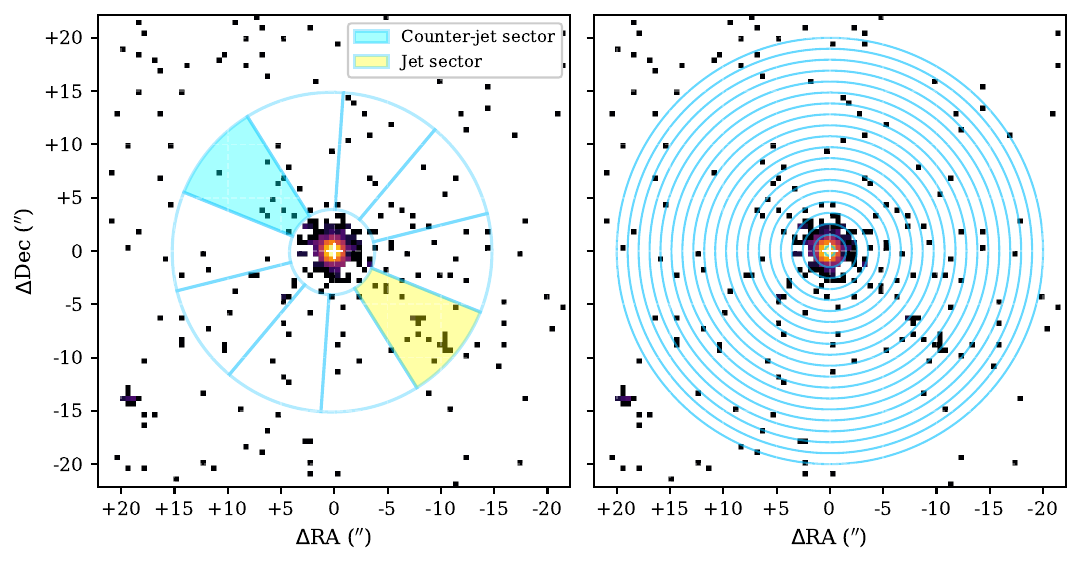}
\caption{An example of the angular (left) and radial (right) sectors used to compute profiles for J1436+6336 merged event files. The highlighted regions in the left plot represent the final extraction regions chosen, and the blue sectors show the sectors used in the sliding window plot.}
\label{fig:comparison_pair}
\end{figure*}

\section{Imaging Analysis}
\label{sec:imaging}
\subsection{PSF}

In order to determine whether the observed extended emission was from the jet itself or from potential asymmetries in Chandra's point spread function (PSF), we generated simulated PSFs using ChaRT \citep{Carter_2003} and MARX v6.0.1 \citep{Davis_2012} for each observation, as given in Table~\ref{tab:wedge_params}. A source spectrum model file was used to weight the simulated rays using the source's spectral energy distribution, as discussed in Section~\ref{sec:core_spec}. Given the faintness of our sources, we left the parameters \texttt{readout\_streak} and \texttt{pileup} to their default values of \texttt{no}. An example PSF model for J1436+6336 is shown in Figure~\ref{fig:real_vs_psf}.

The simulated PSF for each observation was normalized using a circular region $2^{\prime\prime}$ around the center of the core. The function \texttt{dmextract} was then used to find the aperture counts, surface brightness, and errors; we scaled the simulated core PSF fluxes by the ratio of observed to simulated net counts. The final energy range used in our calculations is 0.5--7.0 keV.

\subsection{Radial Profiles}

To identify extended emission around each source, we created radial surface brightness profiles using concentric annuli from the central point of each source. The inner radius was fixed at $0.5^{\prime\prime}$ to exclude the innermost PSF core, and the outer radius was chosen individually for each source, ranging from $11^{\prime\prime}$ to $20^{\prime\prime}$. The surface brightness of the background was measured for each field from $40^{\prime\prime}$ regions, excluding the source AGNs as well as other potential foreground or background sources found through \texttt{wavdetect}.

\begin{figure*}[ht!]
\centering
\begin{subfigure}[b]{0.48\textwidth}
    \centering
    \includegraphics[width=\textwidth]{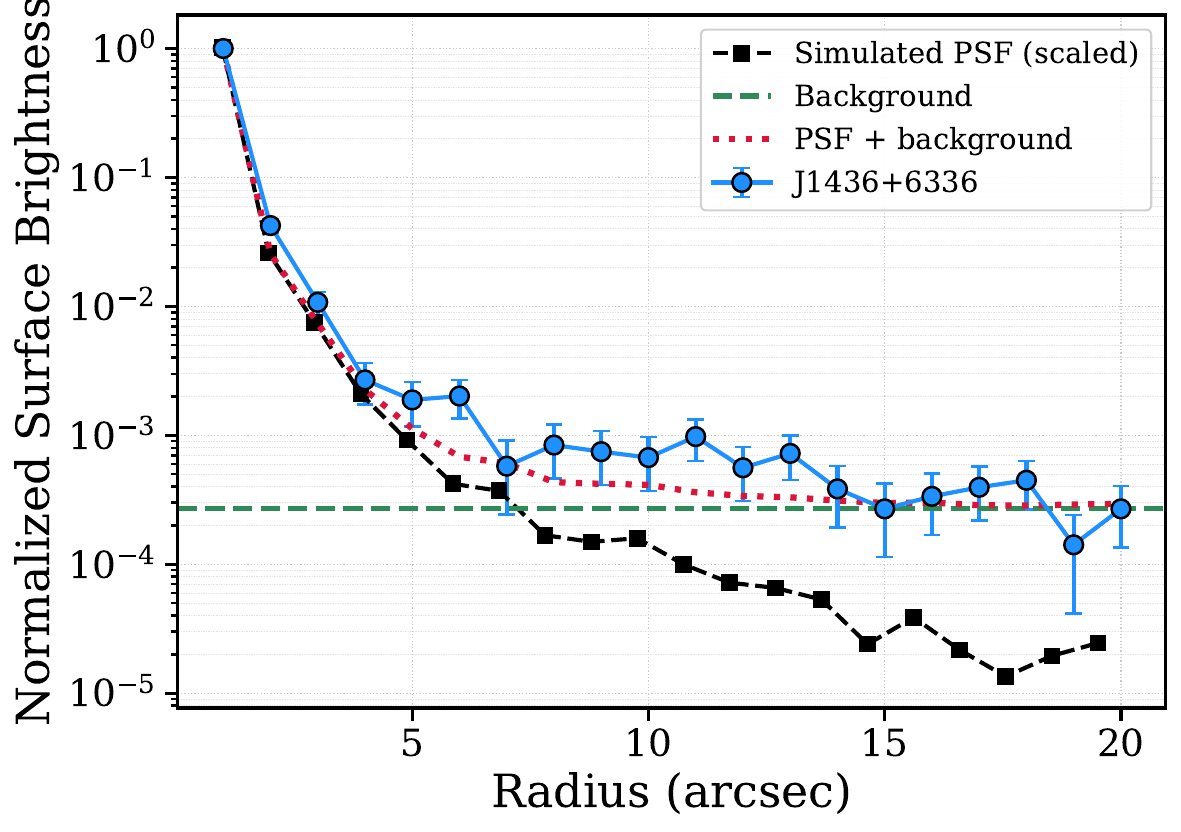}
    \caption{J1436+6336}
    \label{fig:radial8C}
\end{subfigure}
\hfill
\begin{subfigure}[b]{0.48\textwidth}
    \centering
    \includegraphics[width=\textwidth]{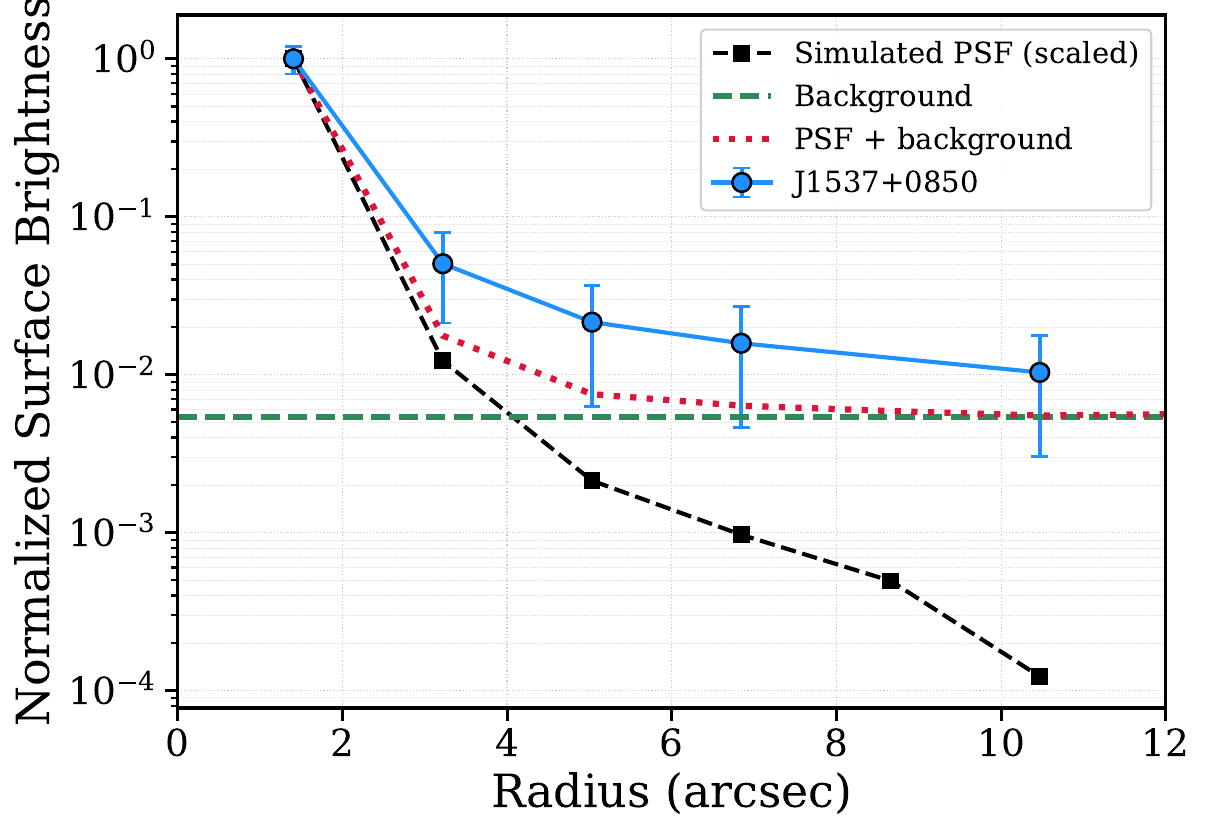}
    \caption{J1537+0850}
    \label{fig:radialSDSS}
\end{subfigure}
\vspace{0.1cm}
\begin{subfigure}[b]{0.48\textwidth}
    \centering
    \includegraphics[width=\textwidth]{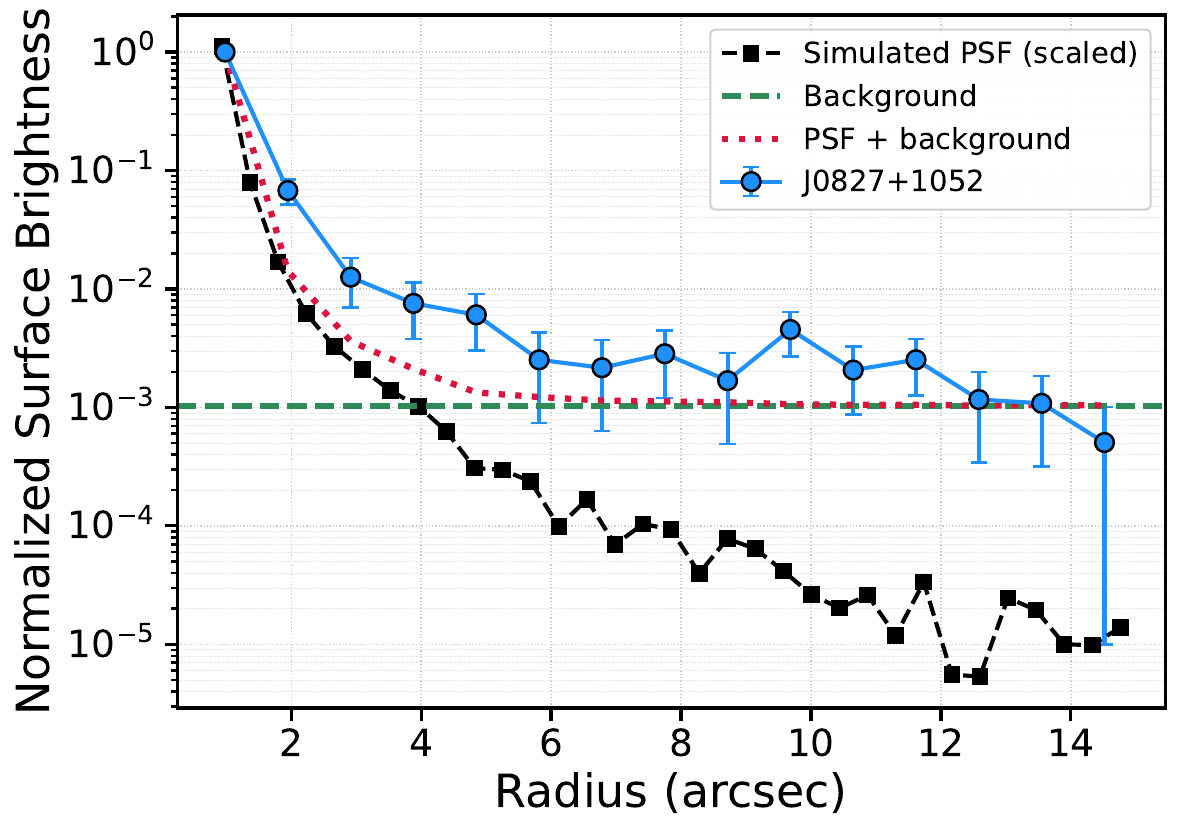}
    \caption{J0827+1052}
    \label{fig:radialB0824}
\end{subfigure}
\hfill
\begin{subfigure}[b]{0.48\textwidth}
    \centering
    \includegraphics[width=\textwidth]{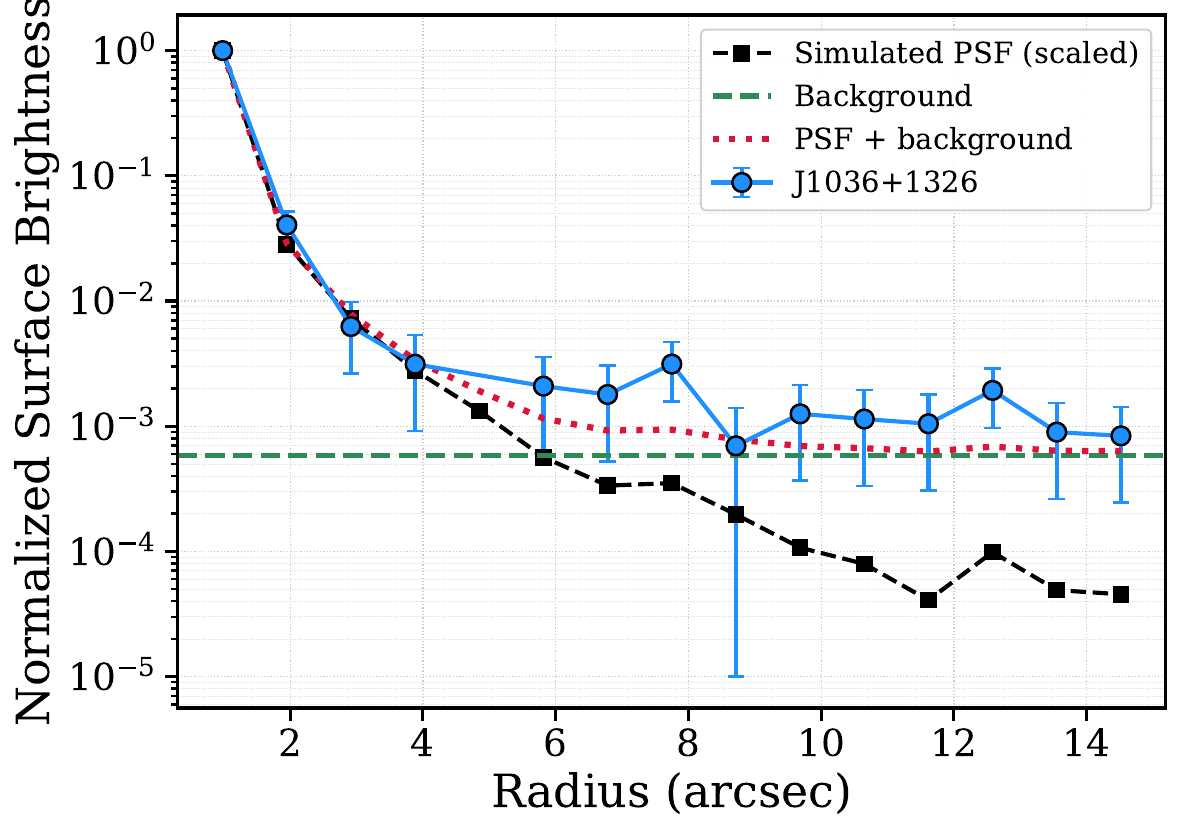}
    \caption{J1036+1326}
    \label{fig:radialB1033}
\end{subfigure}
\caption{The observed normalized surface brightness profiles of the AGNs by radius (blue) compared to the expected profiles of simulated PSFs (black) and local background levels (green). Where the AGN profile exceeds the sum of those two components (red) is an indication of extra emission beyond the AGN core.}
\label{fig:radial_all}
\end{figure*}

Radial profiles comparing observed counts with background and PSF values are shown in Figure \ref{fig:radial_all}. In all four sources, the profiles show emission exceeding the core-normalized PSF and background prediction at large radii. 

\subsection{Annular Wedge}

The jet or counterjet sector for sources is defined a priori as the sector containing the measured kpc-scale radio jet position angle, as estimated using a line region drawn in DS9 along the radio jet position angle, shown in Table \ref{tab:wedge_params}. In this way the sector placement was not decided by the X-ray data itself. Each annulus was divided into equal sectors and measured moving eastward (Figure~\ref{fig:comparison_pair}). The background contribution to each sector is the measured surface brightness multiplied by the sector area.

Table~\ref{tab:detection} shows the significance of the detection of the jet and counterjet regions for AGN samples. In order to investigate potential dependencies that the chosen inner and outer radius values may have on the significance outputs, these values are varied and individual significances are calculated through a grid search in Appendix~\ref{app:grid}; the background annulus is also varied, with grids computed for a 20--35$\arcsec$ background and for a 35--50$\arcsec$ background. The significance of the observed excess was computed using a Poisson survival function, comparing the observed number of counts in the specified sector to fixed expected counts. The expected number of counts in the band are calculated as the sum of the background contribution and the scaled PSF contribution of the ChaRT/MARX runs, with the PSF scale calibrated to the net counts of the AGN core relative to its PSF counts, as shown in Table \ref{tab:detection}. 

\section{Spectral Analysis}
\label{sec:spectral}

\subsection{Core}
\label{sec:core_spec}

We extract spectra using a circular aperture region centered on the AGN core with the CIAO function \texttt{specextract}. We fit the data assuming an absorbed power law model using Sherpa \citep{Siemiginowska_2024}, and in doing so, we limit the spectral energy range to the 0.5--7.0 keV band. The likelihood statistic \texttt{wstat} \citep{Cash_1979, Wachter_1979} was used in order to avoid Gaussian assumptions. We analyzed the spectra of each exposure individually, and combined them for the sources with multiple exposures using \texttt{combine\_spectra}. The Galactic hydrogen column density $N_\mathrm{H}$ was fixed to the value obtained from the \ion{H}{1} map for each source \citep{Bekhti_2016}. The hydrogen column densities, photon indices, and fluxes (0.5--7.0 keV) for each source are detailed in Table \ref{tab:combined_results}. Uncertainties for the fitted parameters, fluxes, and luminosities are given at the 1$\sigma$ level. Rest frame luminosities (2--10 keV) were computed from the fitted model over the observed frame band, shown in Table \ref{tab:combined_results}.

\subsection{Extended Emission}

We extract the extended emission spectra using the jet wedge used in our angular profiles of extended emission. Extractions were performed with \texttt{specextract} on each observation's individual event file; for sources with multiple exposures, the individual spectra were combined using \texttt{combine\_spectra} before fitting. Similar to core extraction, an absorbed power law (\texttt{xsphabs} $\times$ \texttt{powlaw1d}) was fitted to each source using the likelihood of \texttt{wstat} with the simplex optimizer, with the density of the Galactic hydrogen column fixed for each object. Fluxes were computed by integrating the fitted model over the 0.5--7.0 keV band; the fitted photon indices and fluxes are given in Table \ref{tab:combined_results}. While the extraction region may contain a slight PSF contribution, this was not subtracted in the spectral fitting due to the low number of expected counts.

Table \ref{tab:combined_results} summarizes the results of the spectral fit for the jet and counterjet regions of each source. For all regions except the J1436+6336 jet and counterjet, the photon index was fixed at 2.0, and uncertainties are derived from the amplitude at fixed $\Gamma$. All fits converged to a physically constrained normalization. 

\section{Discussion}
\label{sec:discussion}

\subsection{Emission from the core}

\begin{figure}[t]
\centering
    \includegraphics[width=\columnwidth]{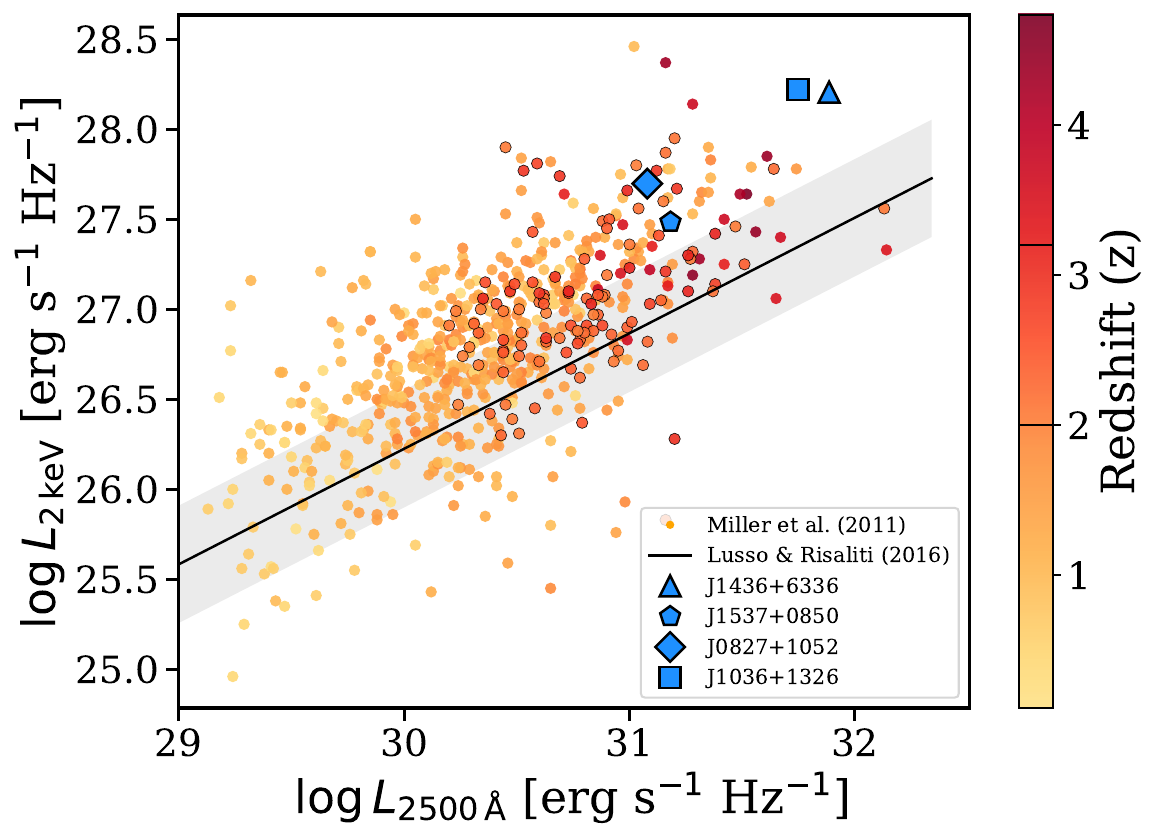} 
    \caption{Rest-frame monochromatic luminosity at 2\,keV versus 2500\,\AA\ for the four observed cores, with \citet{Miller_2011} sources used for comparison. The solid line gives the radio-quiet quasar relation \citep{Lusso_2016}, and the gray band shows intrinsic scatter (0.33 dex).}
    \label{fig:LL}
\end{figure}

To place our values in the context of the literature, we compare the nominal and upper limit values of rest-frame 2\,keV and 2500\,\AA\ core luminosities in Figure~\ref{fig:LL} with a sample of radio-loud quasars from \citet{Miller_2011}. We also show the relation between the 2500\,\AA\ and 2\,keV luminosity from \citet{Lusso_2016} using a sample of radio-quiet quasars with scatter included. The four sources surpass this relation by 0.5--0.9 dex, or 1.5--2.7$\sigma$ taking into account 0.33 dex of intrinsic scatter. When compared against the 106 radio-loud quasars between $2 < z < 3.2$ from \citet{Miller_2011}, the offsets from the \citet{Lusso_2016} relation are in the 90th percentile for J1036+1326, J0827+1052 and J1436+6336 (with offsets $>$0.77 dex), and the $\approx$75th percentile for J1537+0850 with an offset of +0.50 dex. 

The strong X-ray emission in comparison to the UV radiation of these objects, as well as their high radio loudness ($R>100$) and the relatively flat radio spectrum of three of the four core sources ($\alpha_r \approx -0.3$ to $0.3$, with J1537+0850 as an exception with $\alpha_r = 0.89$), points to these sources being blazars. This also aligns with the criteria put forth in \citet{Ighina_2019} to identify a blazar: a relatively flat X-ray spectrum ($\Gamma_X < 1.8$) and strong X-ray emission relative to the optical, $\tilde\alpha_{\rm ox} < 1.355$, where $\tilde\alpha_{\rm ox} = -0.303\,\log(L_{\rm 10\,keV}/L_{2500\,\text{\AA}})$ is the optical-to-X-ray index computed between rest-frame 2500\,\AA\ and 10\,keV. The four observed sources are within this range, with $\tilde\alpha_{\rm ox} = 1.17$--$1.29$. J0827+1052 and J1436+6336 show $\Gamma_X < 1.8$, J1537+0850 is approaching the threshold at $\Gamma_X = 1.78^{+0.17}_{-0.17}$, and J1036+1326 is above the threshold at $\Gamma_X = 2.05^{+0.14}_{-0.14}$. In addition, supplementary data from the MOJAVE survey for J1436+6336 measures a jet viewing angle of $\theta = 9.9^\circ$, as well as a bulk Lorentz factor of $\Gamma = 11.3$ and Doppler factor of $\delta_D = 4.70$, giving direct evidence that the jet direction is near line of sight \citep{Homan_2021}.
Blazar targets are useful when examining extended kpc-scale emission \citep{Meyer_2019}, as their emission is enhanced by Doppler boosting, although this enhancement is mild compared to that of the core. The properties and their categories found here are useful for future reference for choosing observational candidates for kpc-scale jets at high redshift.

\subsection{Emission from the extended jets}

\begin{deluxetable*}{llccc}
\tablewidth{\textwidth}
\tablecaption{X-ray jet emission properties and significance.\label{tab:detection}}
\tablehead{
\colhead{Source} & \colhead{Type} & \colhead{Obs} & \colhead{PSF+bkg} & \colhead{$p$} \\
\colhead{} & \colhead{} & \colhead{(counts)} & \colhead{(counts)} & \colhead{}
}
\startdata
J1436+6336 & J  & 22 & 4.83 & $1.4\times10^{-8}$ \\
J1436+6336 & CJ & 10 & 4.16 & $1.1\times10^{-2}$ \\
J1537+0850 & J  & 5  & 1.89 & 0.045 \\
J1537+0850 & CJ & 2 & 1.86 & 0.56 \\
J0827+1052 & J  & 11 & 1.82 & $4.6\times10^{-6}$ \\
J0827+1052 & CJ & 4 & 2.01 & 0.15 \\
J1036+1326 & J  & 6  & 1.55 & $5.2\times10^{-3}$ \\
J1036+1326 & CJ & 4  & 1.76 & 0.10
\enddata
\tablecomments{Obs is the number of observed 0.5--7.0\,keV counts in the sector containing the jet (J) or counterjet (CJ), defined from the radio contours. PSF+bkg is the expected number of counts in that sector from the scaled point-spread function plus background. $p$ is the Poisson probability of observing at least that many counts from PSF and background alone; we adopt $p<0.05$ as a detection.}
\end{deluxetable*}

We detect significant levels of X-ray extended emission in three mid-redshift AGNs using the Poisson survival function, centered on the region of each jet or counterjet detection using complementary radio contour data. Using a chosen threshold of p = 0.05, J1436+6336 shows a significant detection in the jet region ($p=1.4\times10^{-8}$); a candidate counterjet is also detected in J1436+6336 ($p=1.1\times10^{-2}$). J0827+1052 (J, $p=4.6\times10^{-6}$) and J1036+1326 (J, $p=5.2\times10^{-3}$) both show significant detections, with particular emphasis on the J0827+1052 source, which had 11 observed counts in the flux sector compared to 1.82 expected counts. However, for J1036+1326 (CJ), despite radio contours confirming the presence of extended emission in both FIRST and VLASS, no significant X-ray excess was detected in this observation ($p=0.10$). The counterjet value is reported in Table~\ref{tab:combined_results} although we do not consider this detection to be significant. The J1537+0850 jet region shows marginal significance ($p=0.045$). 

\begin{figure*}
    \centering
    \includegraphics[width=0.95\textwidth]{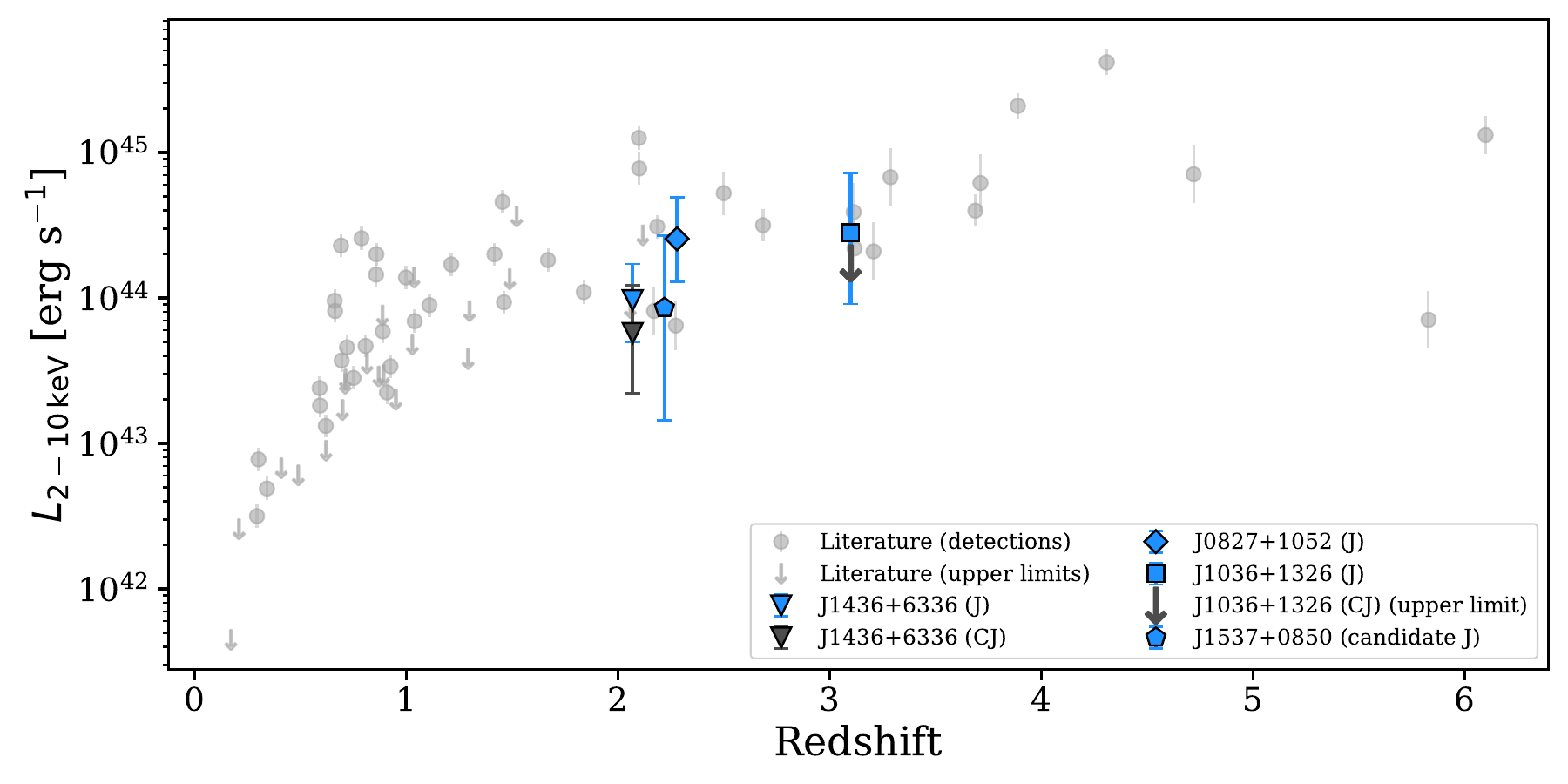}
    \caption{Rest-frame 2--10\,keV X-ray luminosity of kpc-scale jets as a function of redshift. Gray points show previously detected jets (circles) and upper limits (arrows) from a compilation of literature sources as described in the text.}
    \label{fig:jet_z}
\end{figure*}

Figure~\ref{fig:jet_z} compares the X-ray luminosities of our jets with those of the population previously reported in the literature \citep{Siemiginowska_2003, Cheung_2006, McKeough_2016, Simionescu_2016, Marshall_2018, Schwartz_2020, Worrall_2020, Snios_2021, Connor_2021, Ighina_2022}. Our sources lie at $z \approx 2.1$--$3.1$ with $L_{2-10\,{\rm keV}} \approx 9\text{--}28 \times 10^{43}\,{\rm erg\,s^{-1}}$, consistent with previously detected jets at similar redshift, and toward the lower end of the detected population at $z > 2$. The apparent rise in jet luminosity with redshift is largely a selection effect, since only the most luminous jets are detectable at high redshift.

\subsection{SED analysis}

Although a robust analysis of the jet emission mechanisms is beyond the currently available data, we can nevertheless place limiting constraints on the dominant X-ray mechanism by adopting a relatively simple one-zone leptonic model for the IC/CMB emission. We modeled the spectral energy distribution (SED) of source J1436+6336 using \texttt{agnpy} \citep{Nigro_2022}. We defined a spherical emitting region of radius $R_b$ and magnetic field B, with a constant relativistic velocity $\beta c$ and bulk Lorentz factor $\Gamma$ \citep{Nigro_2022}. J1436+6336 was chosen due to additional information available for this source through the MOJAVE survey, including an independently measured viewing angle ($\theta = 9.9^\circ$) \citep{Homan_2021}. \texttt{agnpy.compton} is used to model inverse Compton scattering, with the CMB treated as isotropic, and \texttt{agnpy.synchrotron} is used to model synchrotron emission \citep{Nigro_2022}.

We modeled the electrons as a power law $n_e(\gamma) = k_e\gamma^{-p}$ \citep{Worrall_2009}; we assume values of $p = 2\alpha_{\rm radio}+1 = 3.68$ and $\gamma_{\rm min}=50$. Taking an emitting region of radius $R_b=1.29 \times 10^{22}$\,cm (corresponding to $0\farcs5$), we fit the synchrotron spectrum to the 1.4 and 5\,GHz rest-frame flux densities (Table~\ref{tab:combined_results}), using $\gamma_{\rm max}=8000$ to keep the cutoff above both frequencies and reproduce the slope.

The bulk Lorentz factor of the kpc-scale jet is not known: MOJAVE also provides $\Gamma$ and $\delta_D$ for this source, but these are measured on pc scales and cannot be propagated with accuracy onto the kpc scale. We therefore modeled the IC/CMB scenario over a grid of $\Gamma = 1.5$--$10$, keeping $\theta = 9.9^\circ$ fixed and computing $\delta_D = [\Gamma(1-\beta\cos\theta)]^{-1}$ at each step. For each $\Gamma$, the magnetic field is the only remaining free parameter. For a value of $B$, we fit $k_e$ using a log-space least-squares fit to the radio data. We then compute the IC/CMB flux produced by that same electron population at the observed X-ray frequency. Because the synchrotron flux scales as $k_e B^{(p+1)/2}$ while the IC/CMB flux scales as $k_e$, the predicted X-ray flux decreases with $B$. Therefore, for each value of $\Gamma$ there is a potential $B$ value that reproduces the observed X-ray flux.

We calculate the jet power transported through the emitting region using the equation from
\citet{Ghisellini_2010},
\begin{equation}
    P_{\rm jet} \simeq \pi R_b^2\,\Gamma^2 \beta c\,(U_e + U_B),
    \label{eq:pjet}
\end{equation}
where $U_B = B^2/8\pi$ and $U_e = k_e m_e c^2 (\gamma_{\rm min}^{2-p}-\gamma_{\rm max}^{2-p})/(p-2)$. Across the grid, the jet power is dominated by the magnetic field. It reaches a minimum of $P_{\rm jet} \approx 5\times10^{46}\,{\rm erg\,s^{-1}}$ at $\Gamma \approx 2$ and increases with $\Gamma$, up to $P_{\rm jet} \approx 2\times10^{48}\,{\rm erg\,s^{-1}}$ at $\Gamma = 10$. 

Following the ratio between jet and counterjet flux densities (see eq. 1 in \citet{Spingola_2019}), we find a ratio of $>$340, assuming an intrinsically symmetric Doppler-boosted X-ray emission from either synchrotron or Inverse Compton origin. This estimate is much higher than our observed ratio of 1.84, which could indicate that the jet/counterjet do not share the same pc-scale $\theta$ or bulk velocity. Using the observed ratio to derive the intrinsic velocity of the jet we find $\Gamma\sim1$, suggesting that the observed emission could originate from a lobe-associated structure. Detailed VLBI observations of the extended emission in this target will help understand its morphology \citep[e.g.][]{Spingola_2020}.

A power of $5 \times 10^{46}\,{\rm erg\,s^{-1}}$ is within reasonable limits, and therefore we cannot exclude IC/CMB as a potential dominant emission mechanism for the X-ray emission in J1436+6336. Future modeling work will be improved with deeper radio observations at sub-arcsecond resolution with the VLA and e-MERLIN, which will be able to better resolve the kpc-scale jet structure, allowing for more realistic modeling of the initial electron population in the radio band (see also \citealt{Spingola_2020}). Additional Chandra X-ray observations of these objects with longer exposures would tighten the constraints on the jets and on the X-ray photon index. The value of deeper X-ray data has already been shown using the $z = 6.1$ quasar PSO J0309+27: after an initial jet detection by \cite{Ighina_2022}, \cite{Meanti_2026} used 22 Chandra observations to resolve two distinct features of the jet and constrain its spectral properties. Improved radio and X-ray observations would allow us to better constrain the dominant X-ray emission mechanism in mid-redshift jets.

\section{Summary}

In this work, we have presented an analysis of Chandra X-ray observations of four radio-loud AGNs at $z \approx 2.1$--$3.1$, each with a radio jet detected at parsec scales in previous VLBI observations. After correcting the astrometry of each observation and modeling the PSF of each quasar core, we detect significant extended X-ray emission on kpc scales in three of the four sources, with marginal emission in J1537+0850, using a Poisson survival function evaluated in angular wedge regions. The extended X-ray emission is detected along the direction of both the VLBI-detected parsec-scale jet and the extended emission in lower-resolution VLA imaging. The discovery of these X-ray jets brings the number of X-ray-detected kpc-scale jets at $z > 2$ to 21. This builds toward the sample needed for population studies of high-redshift jet emission.

We identified a candidate X-ray counter-jet in J1436+6336 ($p=0.011$). Deeper observations are needed to confirm this detection. 
Using spectral fitting of the four sources, we determined the X-ray fluxes, luminosities, and photon indices for the jets. Combining X-ray measurements with radio data, we show that IC/CMB can feasibly reproduce the observed X-ray flux for J1436+6336 without requiring extreme jet powers ($P_{\rm jet} \sim 5\times10^{46}\,{\rm erg\,s^{-1}}$ for $\Gamma = 2$). With a single X-ray flux point, we cannot yet distinguish between the two mechanisms. A more useful test is to model both scenarios in terms of the required jet power, magnetic field and bulk Lorentz factor; however, both scenarios require more rigorous modeling before conclusions can be drawn. 

Future modeling attempts would greatly benefit from deeper radio observations at sub-arcsecond resolution using VLA and e-MERLIN. Longer Chandra exposures would also allow us to better constrain the X-ray jet morphology along with the calculated photon flux and luminosity values. Optical and ultraviolet observations, such as broad-band UV photometry to measure jet knot flux densities, would also provide additional information to further constrain spectral shapes between the radio and X-ray bands.

The IC/CMB model predicts that as the CMB energy density increases with $(1+z)^4$, the ratio of X-ray to radio emission in jets should increase with redshift. As the sample of X-ray jets at $z > 2$ increases, it will become possible to test this prediction and to determine which emission mechanism dominates jet emission in the early Universe.

\begin{acknowledgements}
{\small
We want to thank A. Caccianiga for helpful discussions. 

The scientific results reported in this article are based on observations made by the \textit{Chandra X-ray Observatory} contained in~\dataset[CDC 730]{https://doi.org/10.25574/cdc.730}. This research has made use of software provided by the \textit{Chandra} X-ray Center (CXC) in the application package CIAO. The National Radio Astronomy Observatory and Green Bank Observatory are facilities of the U.S. National Science Foundation operated under cooperative agreement by Associated Universities, Inc. 

Support for this work was provided by the National Aeronautics and Space Administration through Chandra Award Numbers GO5-26066X and GO3-24069X issued by the Chandra X-ray Center, which is operated by the Smithsonian Astrophysical Observatory for and on behalf of the National Aeronautics Space Administration under contract NAS8-03060. 

K.B. acknowledges proofreading the text of this paper with assistance from Claude Opus 5.5 (Anthropic). T.C. and A.S. acknowledge support from NASA Contract NAS8-03060 to the Chandra X-ray Center. C.~S. acknowledges financial support from INAF $-$ Ricerca Fondamentale 2024 (Ob.~Fu. 1.05.24.07.04) and by the Italian Ministry of University and Research (grant FIS2023$-$01611, CUP C53C25000300001, \textsl{DARKER: Accurate constraints on dark matter and dark energy in the era of Precision Cosmology}). \textcolor{black}{This work was supported by the INAF $-$ Astrofisica di Frontiera 2025, Excellence Track grant "PARSE" (PI Vanzella).}  
}
\end{acknowledgements}

\facilities{CXO (ACIS), VLA, VLBA}
\software{agnpy \citep{Nigro_2022}, ChaRT \citep{Carter_2003}, CIAO \citep{Fruscione_2006, Fruscione_2026}, MARX \citep{Davis_2012}, SAOImageDS9 \citep{Fruscione_2026_SAODS9}, Sherpa \citep{Siemiginowska_2024}}

\appendix
\onecolumngrid
\section{Significance Maps}
\label{app:maps}

\begin{figure*}[ht!]
\centering
\begin{minipage}{0.46\textwidth}
    \centering
    \includegraphics[width=\linewidth]{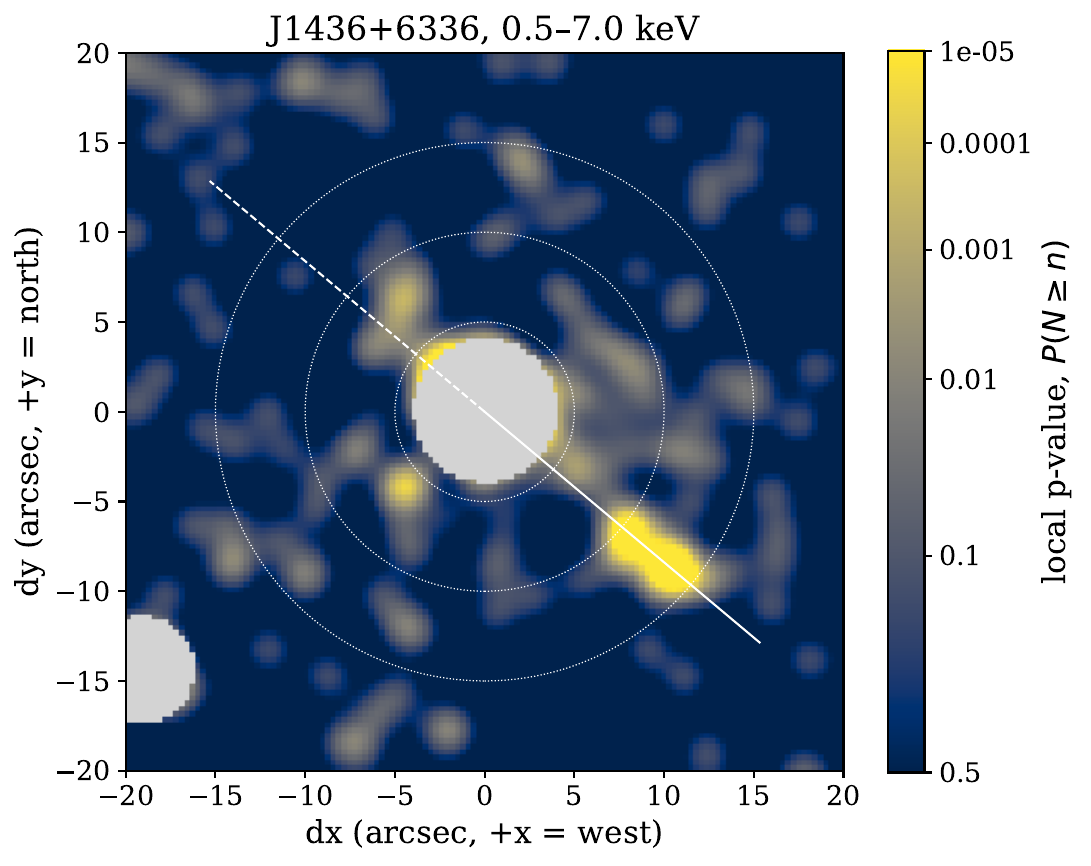}
\end{minipage}
\hfill
\begin{minipage}{0.46\textwidth}
    \centering
    \includegraphics[width=\linewidth]{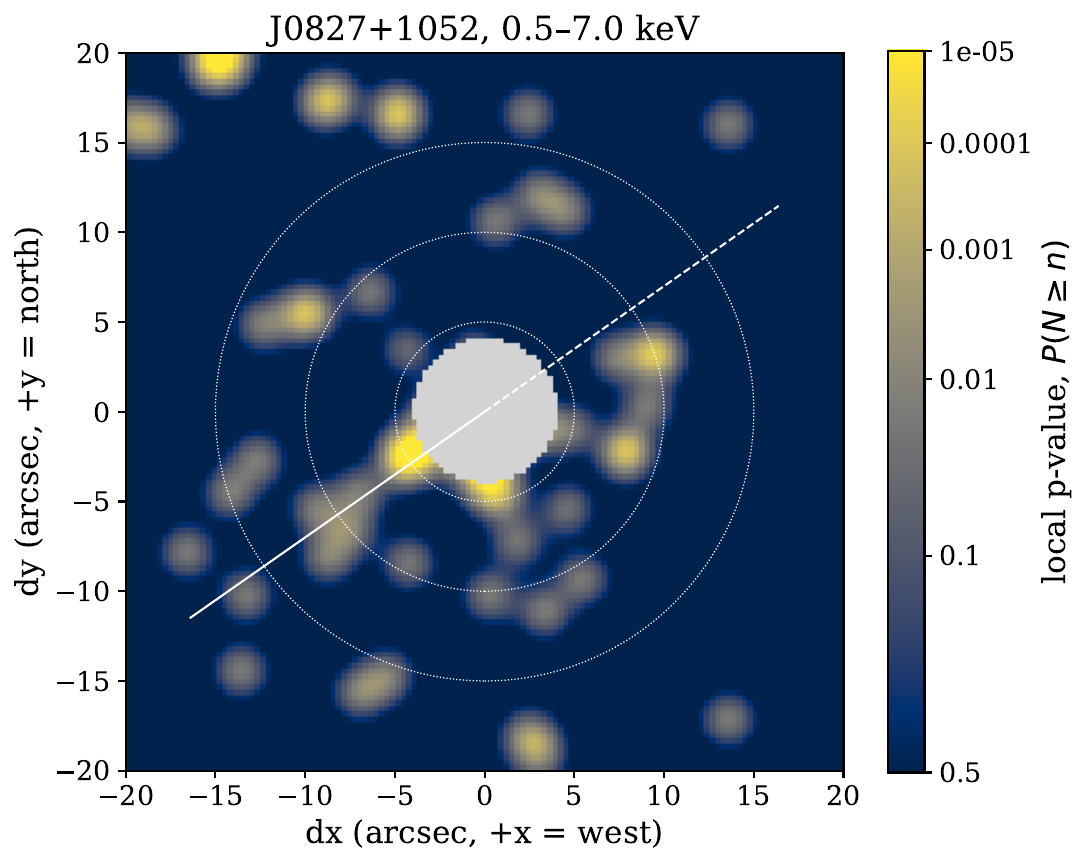}
\end{minipage}

\vspace{2mm}

\begin{minipage}{0.44\textwidth}
    \centering
    \includegraphics[width=\linewidth]{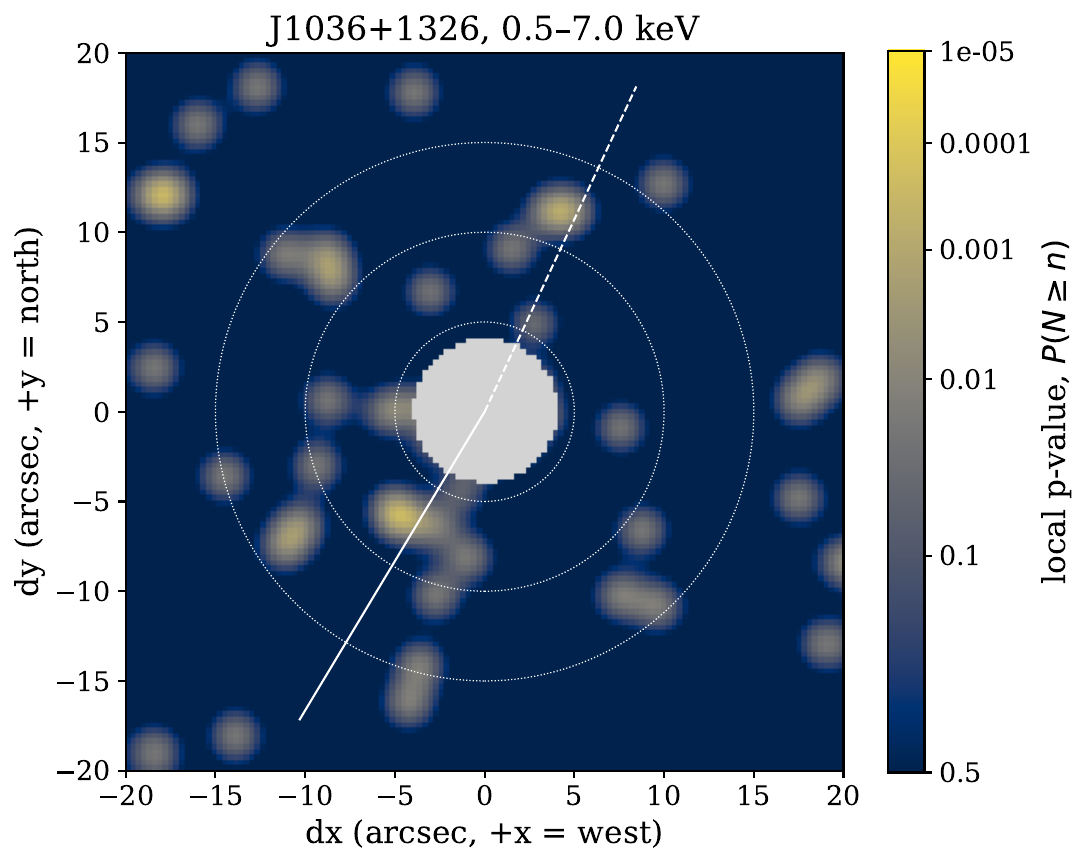}
\end{minipage}
\hfill
\begin{minipage}{0.44\textwidth}
    \centering
    \includegraphics[width=\linewidth]{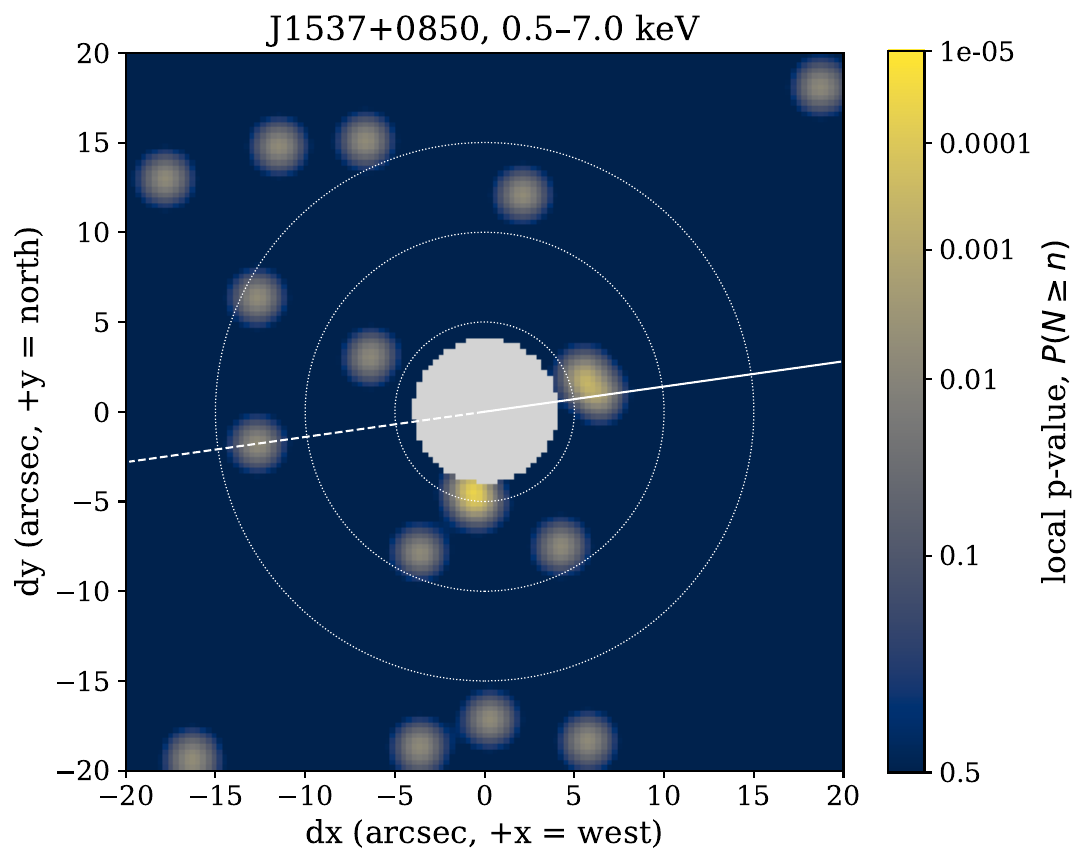}
\end{minipage}
\caption{Broad-band (0.5--7.0~keV) excess-significance maps after subtracting the PSF model and background. The color shows the significance in $\sigma$; the gray regions mask the core ($r < 4\arcsec$) and unrelated field sources.}
\label{fig:appendixA}
\end{figure*}

We constructed excess significance maps for each source in order to clearly visualize and isolate the extended X-ray emission from the quasar core (Fig \ref{fig:appendixA}). We simulated the Chandra core PSF using ChaRT and MARX, and then normalized the simulation to the observed counts. We measured the background surface brightness in a 35--50$\arcsec$ annulus, with a mask around unrelated field sources, and core-anchored the PSF to account for potential wings or artifacts. We masked the core for a radius less than 4$\arcsec$ along with nearby field sources. We binned the observed events and the model (PSF + background) onto 0.3$\arcsec$ pixels in the 0.5--7.0 keV band, summed over all observations of each source. We smoothed both with a Gaussian kernel of $\sigma = 1\arcsec$.

\section{Significance Grid}
\label{app:grid}

\begin{figure*}[ht!]
    \centering
    \includegraphics[width=\linewidth]{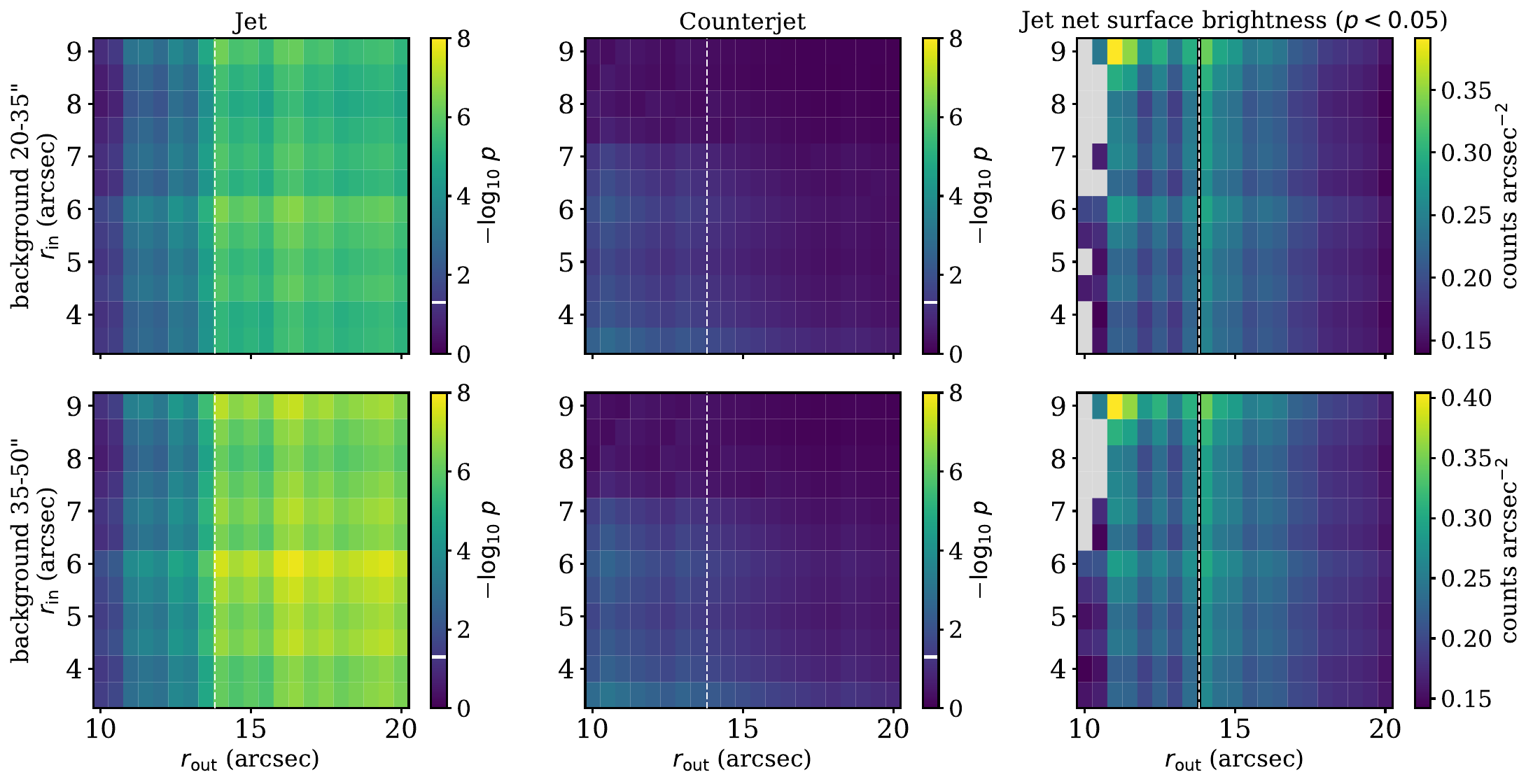}
    \caption{A characterization of the effects of systematic choices on the derived jet significance for J1436+6336. Each cell represents a wedge along the radio jet axis, with varying inner and outer radii. Rows represent different background values (20--35\arcsec top and 35--50\arcsec bottom). The right panels show the net surface brightness of the jets; points are colored gray at p < 0.05 Dashed lines represent the distance of the J1436+6336 jet knot.}
    \label{fig:excess_soft}
\end{figure*}

To support our significance reports, we tested whether the jet detection depends on the choice of extraction region for J1436+6336 (Fig \ref{fig:excess_soft}). We repeated the wedge measurement over a grid of inner and outer radii, $r_{\rm in} = 3.5$--$9\arcsec$ and $r_{\rm out} = 10$--$20\arcsec$ in steps of $0.5\arcsec$, for a 36\textdegree\ wedge centered on the radio jet axis. We also varied the grids over two background annuli, 20--35$\arcsec$ and 35--50$\arcsec$, respectively. We found that the significance value is dependent on the outer annulus and aligns with the distance of the photon excess at approximately 10--15$\arcsec$.

{\setlength{\tabcolsep}{3pt}
\movetabledown=55mm
\begin{rotatetable*}
\begin{deluxetable*}{llccccccccccc}
\tabletypesize{\scriptsize}
\tablecaption{Radio and X-ray properties for the AGN samples.\label{tab:combined_results}}
\tablehead{
\colhead{Source} & \colhead{Comp.} & \colhead{$N_\mathrm{H}$} & \colhead{$\Gamma$} & \colhead{Ampl.} & \colhead{$F_{0.5\text{--}7}$keV} & \colhead{$L_{2\text{--}10}$keV} & \colhead{$L_\nu(4400\,\text{\AA})$} & \colhead{$L_\nu(2500\,\text{\AA})$} & \colhead{$\alpha_{\rm radio}$} & \colhead{$L_{1.4\,{\rm GHz}}$} & \colhead{$L_\nu(5\,{\rm GHz})$} & \colhead{$R$} \\
\colhead{} & \colhead{} & \colhead{($10^{20}\,\mathrm{cm^{-2}}$)} & \colhead{} & \colhead{($\mathrm{cm^{-2}\,s^{-1}\,keV^{-1}}$)} & \colhead{($\mathrm{erg\,cm^{-2}\,s^{-1}}$)} & \colhead{($\mathrm{erg\,s^{-1}}$)} & \multicolumn{2}{c}{($10^{31}$\,erg\,s$^{-1}$\,Hz$^{-1}$)} & \colhead{} & \multicolumn{2}{c}{($10^{33}$\,erg\,s$^{-1}$\,Hz$^{-1}$)} & \colhead{}
}

\startdata
J1436+6336 & Core & 1.36 & $1.72^{+0.04}_{-0.04}$ & $1.74^{+0.07}_{-0.07}\times10^{-4}$ & $8.98^{+0.44}_{-0.44}\times10^{-13}$ & $1.58^{+0.07}_{-0.07}\times10^{46}$ & $9.85$ & $7.68$ & $-0.27\pm0.15$ & $65\pm11$ & $91.2\pm5.0$ & $930\pm50$ \\
 & J  & & $1.33^{+0.55}_{-0.54}$ & $8.59^{+6.01}_{-3.85}\times10^{-7}$ & $6.26^{+4.38}_{-2.81}\times10^{-15}$ & $9.38^{+7.24}_{-4.55}\times10^{43}$ & \nodata & \nodata & $1.34\pm0.15$ & $24.1\pm4.2$ & $4.40\pm0.24$ & \nodata \\
 & CJ & & $1.53^{+0.83}_{-0.78}$ & $5.60^{+5.87}_{-3.22}\times10^{-7}$ & $3.40^{+3.56}_{-1.95}\times10^{-15}$ & $5.57^{+6.23}_{-3.41}\times10^{43}$ & \nodata & \nodata & \nodata & \nodata & \nodata & \nodata \\
\hline
J1537+0850 & Core & 3.14 & $1.78^{+0.17}_{-0.17}$ & $2.73^{+0.49}_{-0.43}\times10^{-5}$ & $1.35^{+0.31}_{-0.26}\times10^{-13}$ & $2.85^{+0.55}_{-0.48}\times10^{45}$ & $1.95$ & $1.52$ & $0.89\pm0.15$ & $20.3\pm3.7$ & $6.57\pm0.34$ & $338\pm18$ \\
 & J  & & 2.0\tablenotemark{a} & $9.11^{+19.45}_{-7.58}\times10^{-7}$ & $3.85^{+8.23}_{-3.20}\times10^{-15}$ & $8.75^{+18.69}_{-7.28}\times10^{43}$ & \nodata & \nodata & $0.97\pm0.15$ & $16.1\pm2.9$ & $4.67\pm0.25$ & \nodata \\
\hline
J0827+1052 & Core & 2.61 & $1.70^{+0.13}_{-0.13}$ & $4.34^{+0.57}_{-0.52}\times10^{-5}$ & $2.30^{+0.35}_{-0.35}\times10^{-13}$ & $4.99^{+0.72}_{-0.65}\times10^{45}$ & $1.53$ & $1.20$ & $0.23\pm0.15$ & $29.3\pm5.3$ & $22.0\pm1.1$ & $1433\pm74$ \\
 & J  & & 2.0\tablenotemark{a} & $2.44^{+2.27}_{-1.21}\times10^{-6}$ & $1.03^{+0.96}_{-0.51}\times10^{-14}$ & $2.50^{+2.32}_{-1.24}\times10^{44}$ & \nodata & \nodata & $1.10\pm0.17$ & $9.1\pm1.9$ & $2.24\pm0.12$ & \nodata \\
\hline
J1036+1326 & Core & 4.27 & $2.05^{+0.14}_{-0.14}$ & $5.77^{+0.74}_{-0.68}\times10^{-5}$ & $2.37^{+0.40}_{-0.35}\times10^{-13}$ & $1.24^{+0.19}_{-0.17}\times10^{46}$ & $7.17$ & $5.59$ & $0.29\pm0.15$ & $16.1\pm3.5$ & $11.2\pm0.6$ & $156\pm9$ \\
 & J  & & 2.0\tablenotemark{a} & $1.30^{+2.03}_{-0.88}\times10^{-6}$ & $5.50^{+8.57}_{-3.70}\times10^{-15}$ & $2.81^{+4.39}_{-1.90}\times10^{44}$ & \nodata & \nodata & $0.90\pm0.16$ & $22.8\pm5.3$ & $7.2\pm0.4$ & \nodata \\
 & CJ\tablenotemark{c} & & 2.0\tablenotemark{a} & $3.50^{+4.51}_{-2.98}\times10^{-7}$ & $1.48^{+1.91}_{-1.26}\times10^{-15}$ & $7.55^{+9.73}_{-6.43}\times10^{43}$ & \nodata & \nodata & $0.90\pm0.16$\tablenotemark{b} & $9.2\pm2.1$ & $2.9\pm0.2$ & \nodata
\enddata
\tablenotetext{a}{Photon index fixed at $\Gamma=2.0$; luminosity. Uncertainties are from the amplitude at fixed $\Gamma$.}
\tablenotetext{b}{Assumed equal to the jet spectral index.}
\tablenotetext{c}{Not a significant X-ray detection ($p=0.10$; Table~\ref{tab:detection}).}
\tablecomments{X-ray values use power-law fits with Galactic absorption $N_H$ for J1436+6336, J1537+0850, J0827+1052, and J1036+1326. Ampl.\ is the power-law normalization at 1\,keV. $F_{0.5\text{--}7}$ is the unabsorbed observed frame 0.5--7.0\,keV flux, and $L_{2\text{--}10}$ is the rest-frame 2--10\,keV luminosity. All uncertainties are 1$\sigma$. The flux and luminosity uncertainties for the core, J1436+6336 jet, and J1436+6336 counterjet are propagated using $\Gamma$ and amplitude uncertainties. The jet and counterjet flux uncertainties are propagated using a 1$\sigma$ interval on the amplitude at fixed $\Gamma$. Radio calibrations have flux calibration uncertainties of 5$\%$ for FIRST and 10$\%$ from VLASS. $\alpha_{\rm radio}$ is the radio spectral index ($S_\nu\propto\nu^{-\alpha}$) from flux densities at 1.4 and 3\,GHz; rest-frame radio luminosities are K-corrected using $\alpha_{\rm radio}$, and the 5\,GHz values are extrapolated from 1.4\,GHz. The radio loudness is $R = L_\nu(5\,{\rm GHz})/L_\nu(4400\,\text{\AA})$ for each core.}
\end{deluxetable*}
\end{rotatetable*}
}

\bibliography{sample701}{}
\bibliographystyle{aasjournalv7}
\end{document}